\documentclass[prl, twocolumn, showpacs, superscriptaddress]{revtex4-1}
\usepackage{amsmath}
\usepackage{graphicx}
\usepackage{xcolor}
\begin{document}
\title{Surface sulci in squeezed soft solids}
\author{T.\ Tallinen}
\affiliation{School of Engineering and Applied Sciences, Harvard University, Cambridge, MA 02138, USA}
\affiliation{Department of Physics, University of Jyv\"{a}skyl\"{a}, P.O.Box 35, FI-40014 Jyv\"{a}skyl\"{a}, Finland}
\author{J.\ S.\ Biggins}
\affiliation{School of Engineering and Applied Sciences, Harvard University, Cambridge, MA 02138, USA}
\author{L.\ Mahadevan}
\affiliation{School of Engineering and Applied Sciences, Harvard University, Cambridge, MA 02138, USA}
\affiliation{Department of Physics, Harvard University, Cambridge, MA 02138, USA}
\date{\today}
\begin{abstract}
The squeezing of soft solids, the constrained growth of biological tissues, and the swelling of soft elastic solids such as gels can generate large compressive stresses at their surfaces. This causes the otherwise smooth surface of such a solid to becomes unstable when its stress exceeds a critical value. Previous analyses of the surface instability have assumed two-dimensional plane-strain conditions, but in experiments isotropic stresses often lead to complex three-dimensional sulcification patterns. Here we show how such diverse morphologies arise by numerically modeling the lateral compression of a rigidly clamped elastic layer. For incompressible solids, close to the instability threshold, sulci appear as I-shaped lines aligned orthogonally with their neighbors; at higher compressions they are Y-shaped and prefer a hexagonal arrangement. In contrast, highly compressible solids when squeezed show only one sulcified phase characterized by a hexagonal sulcus network.
\end{abstract}
\pacs{ 46.15.-x,  46.32.+x}
\maketitle

Complex patterns often arise from simple causes, in such instances as the fractal structures in physical aggregation phenomena  \cite{witten} or the labyrinthine structures, spots and stripes in chemical systems \cite{turing}. In purely mechanical systems, there are two basic instabilities associated with the buckling of a slender filament or sheet \cite{euler,antman} and the cracking of a bulk solid \cite{griffith,broberg}. The first instability arises because of a competition between compression and bending and is embodied in the ratio of two length scales, while the second arises because of a competition between bulk and surface effects and embodied in the ratio of two energy scales.  Here we show that an even simpler system --- a thick isotropic, homogeneous elastic layer which is subject to planar compression and whose top surface is free -- is susceptible to the formation of various sulcal patterns (cusped folds) as a function of the applied compressive strain. This process is based on a surface instability with no length scale, in sharp contrast to the thin film based pattern formation \cite{bowden} where the film thickness is the intrinsic length scale.

The elastic instability of a compressed surface was first studied by Biot \cite{biot} who showed that a half-space of incompressible neo-Hookean material becomes unstable to any smooth perturbation when the surface stretch ratios $\lambda_x$ and $\lambda_y$  (defined so that an uncompressed surface has $\lambda_x=\lambda_y =1$) reach a critical value $\lambda_x \sqrt{\lambda_y} \approx 0.544$. More recent studies have found that a subcritical instability in the form of a sulcus is energetically favorable when $\lambda_x \sqrt{\lambda_y} \lesssim 0.647$ \cite{hohlfeld}, i.e.\ a nonlinearly unstable sulcified state with no nucleation threshold can exist at lower compression than predicted by Biot's linear analysis. Apart from some early and preliminary studies in gels \cite{tanaka, suematsu}, theoretical and numerical studies of sulcification \cite{hohlfeld, dervaux, jin, mora, cao, dervaux_rev} assume two-dimensional plane-strain conditions, and hence exclude all realistic sulcal morphologies that involve three-dimensional deformations seen in most physical experiments \cite{trujillo, yoon}. These three-dimensional morphologies are also seen in biological tissues, including particularly prominent sulci in the primate brain \cite{welker}, but also in tumors \cite{dervaux} and other organs, where there is evidence of mechanical forces driving folding \cite{xu}. In addition to being a new paradigm for mechanical pattern formation, this system also serves as a model for unusual thermodynamical phase transitions without a barrier \cite{hohlfeld}, classical nucleation \cite{chen}, interface instabilities etc. 

\begin{figure*}[t]
\begin{center}
\includegraphics[width = 140mm]{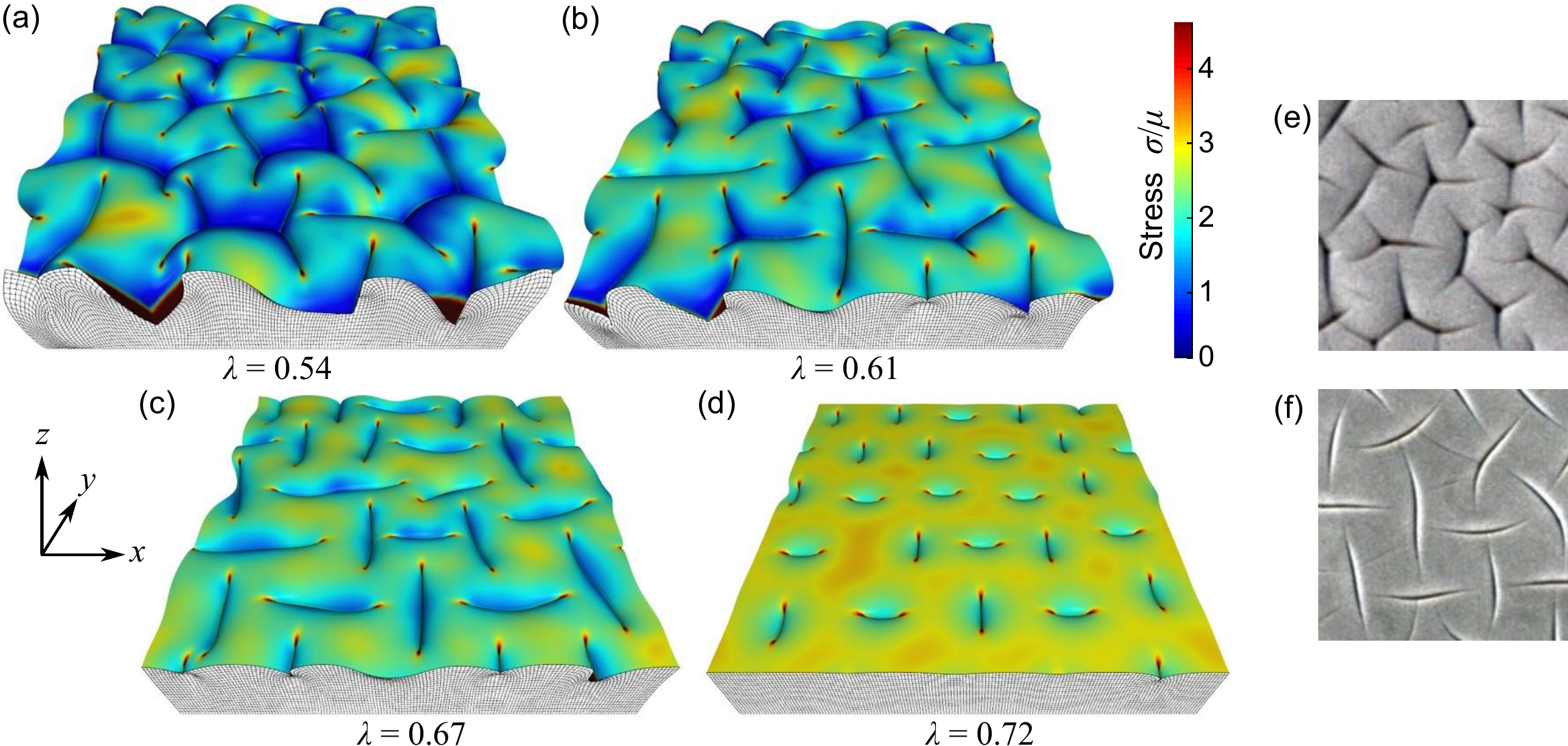}
\end{center}
\caption{
Simulated unfolding of a compressed and sulcified solid layer starts from a state with a transversely isotropic stretch $\lambda = 0.54$ (a). The layer is then decompressed quasistatically and intermediate states are shown for $\lambda = 0.61$ (b), $\lambda = 0.67$ (c) corresponding to the Biot threshold, and $\lambda = 0.72$ (d) just before unsulcification. Coloring indicates the largest compressive stress at the surface. Experimental sulcal patterns are shown in a swelling gel in (e) and (f), courtesy of J.\ Yoon \& R.\ C.\ Hayward.
}
\label{fig1}
\end{figure*}

Experimental observations of sulcal patterns can be realized by isotropic compression induced, for example, by swelling gels \cite{trujillo, yoon, pkim} and typically lead to three-dimensional patterns that are hard to control. Numerical simulations provide means for controlled studies, but simulating sulcification in three dimensions is challenging as it involves finite deformations, tracking the free boundaries of the unknown sulci, and accounting for self-contact of the free surface. To overcome these difficulties, we use a finite element method to approximate the solid with a dense rectangular mesh of tetrahedrons \cite{sm}, based on a discretized finite strain elasticity theory with a neo-Hookean energy density
\begin{equation} \label{w}
W = \frac{\mu}{2} \left[ \textrm{Tr}(\mathbf{F} \mathbf{F}^{\textrm{T}})J^{-2/3} - 3 \right] + K ( J - \log J-1 ),
\end{equation}
where $\mathbf{F}$ is the deformation gradient, $J = \textrm{det}(\mathbf{F})$, and $\mu$ and $K$ are shear and bulk modulus, respectively. To relax the integrated energy density (\ref{w}) towards minima, we  use damped Newtonian dynamics of the nodal degrees of freedom. Self-avoidance is implemented by short range repulsive contacts between the edges and faces that make up the surface. The numerical model is well suited for sulcification studies as Newtonian dynamics naturally allows the system to change rapidly when there are large configuration changes while the quadratic elastic potentials allow for stable integration of the equations of motion with a fixed time step. To confirm that our results are independent of the mesh geometry, we also used a hexagonal prism mesh and checked that our results are robust \cite{sm}.

To study sulcification under isotropic compression, we simulate a layer with thickness $h$ in the stress-free state, and lateral dimensions corresponding to a square of side  $L = 10h$; the lattice has  $\sim$ 320 $\times$ 320 $\times$ 40 nodes with $\approx 2 \times 10^7$ elements. The base of the layer is clamped and periodic boundary conditions are applied along the lateral edges to prevent edge effects from constraining sulcal morpologies. Since many soft materials are approximately incompressible, we assume that the bulk modulus $K = 30\mu$. Our simulations start from an isotropic compressed reference state with stretch $\lambda = \lambda_x = \lambda_y = 0.54$ in both planar directions, corresponding to strain $\epsilon = \lambda-1 = -0.46$ and $\lambda_x \sqrt{\lambda_y} = 0.40$ well beyond the Biot-point ($\lambda_x \sqrt{\lambda_y} \approx 0.544$ \cite{biot}) and T-point ($\lambda_x \sqrt{\lambda_y} \approx 0.647$ \cite{hohlfeld}). To trigger sulcification the featureless flat surface is perturbed by small random vertical displacements of the nodes (maximum amplitude $10^{-3} \times$ lattice constant), after which the system is allowed to relax to a sulcified state while keeping the strain constant. In Fig.\ \ref{fig1}a we see the appearance of a densely sulcified state characterized by isolated Y-shaped triple-junctions of sulci lying on an approximately triangular lattice. The three-fold symmetry of the junctions is consistent with their angle being $\approx 120^{\circ}$; occasionally  some triple-junctions share an arm with their neighbor, although usually they are isolated.  Once the layer is fully relaxed in its equi-biaxially compressed state, we quasistatically decompress it while preserving the lateral stress isotropy (see movie \cite{sm}). Decompression of the layer transforms the Y-shaped sulci, one by one, to I-shaped sulci. In Fig.\ \ref{fig1}b we show the layer at $\lambda = 0.61$ where Y's and I's coexist. At $\lambda = 0.67$ (Fig.\ \ref{fig1}c), at the Biot point, all Y's have transformed to I's, but their number remains approximately constant. As the layer is allowed to relax still more, the I's shorten and eventually unfold at $\lambda \approx 0.73$ as the T-point is approached (Fig.\ \ref{fig1}d). We use unloading rather than loading to probe the patterns to circumvent that the absence of non-smooth perturbations prevents sulcus formation during loading until the Biot-point \cite{biot} is reached, in contrast with any physical experiment where the T-point determines sulcus formation \cite{hohlfeld}.

\begin{figure}[t]
\begin{center}
\includegraphics[width = 82mm]{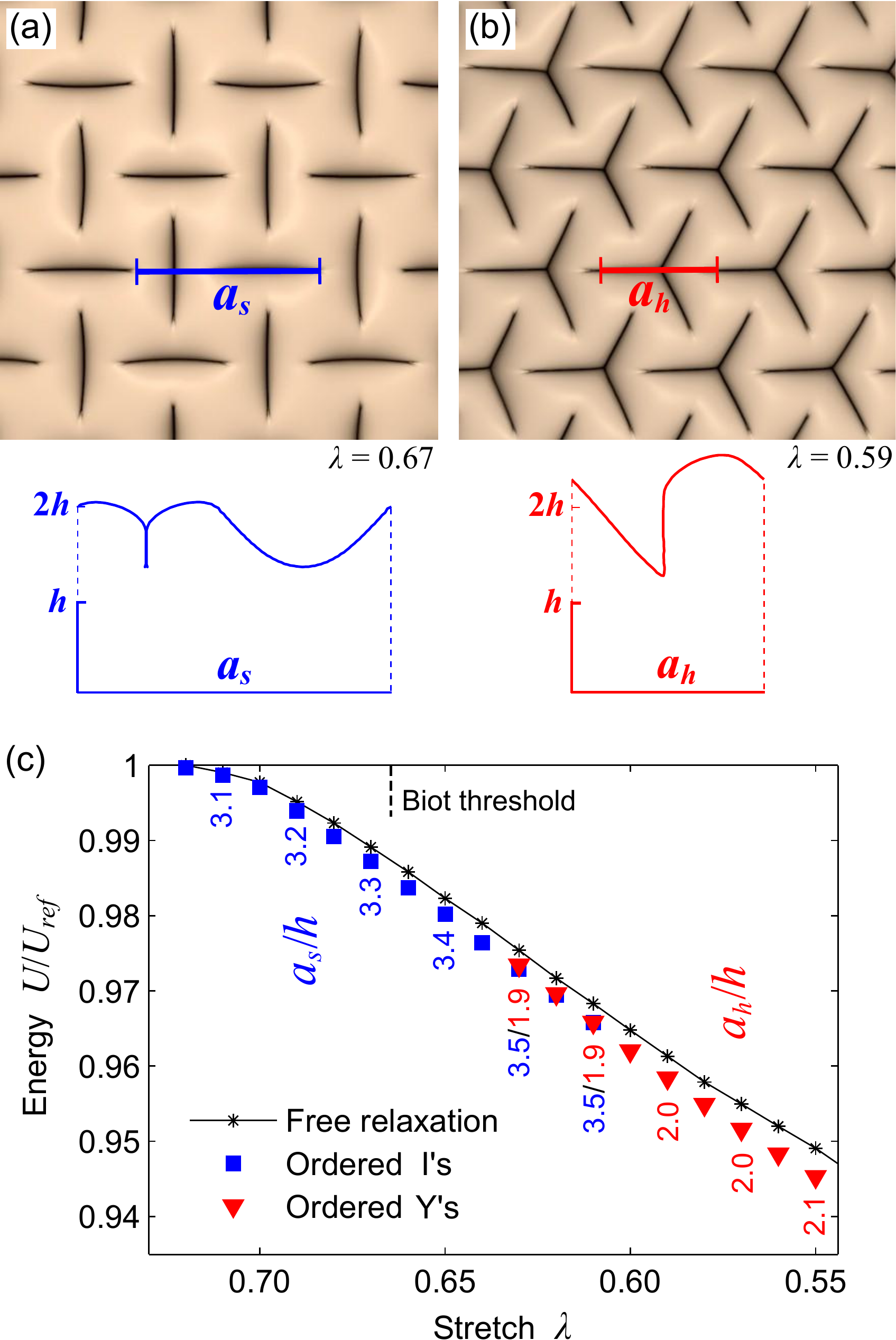}
\end{center}
\caption{
A square lattice (a) of I-shaped sulci with alternating orientation minimizes energy near the sulcification threshold, whereas at higher compression hexagonal arrangement (b) of Y-shaped sulci is favorable. Surface profiles along the lines indicated in (a) and (b) are shown below the patterns.
In (c) the energies of these patterns are compared to that of a freely relaxing layer of Fig.\ \ref{fig1} as a function of stretch $\lambda = \lambda_x = \lambda_y$. All energies are normalized by the energy $U_{ref}$ of an unsulcified reference state. Numbers below the points indicate the optimal spacing of the square (blue) or hexagonal (red) lattice. }
\label{fig2}
\end{figure}

\begin{figure}[t]
\begin{center}
\includegraphics[width = 82mm]{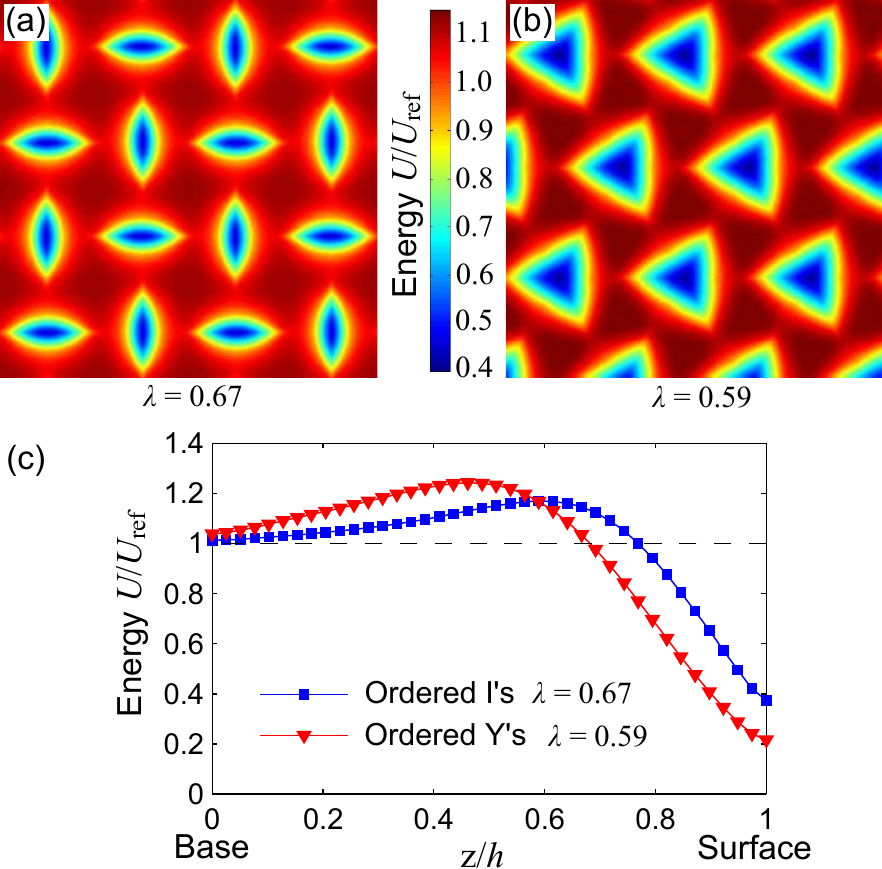}
\end{center}
\caption{
Depth averaged energy distribution for (a) an I-pattern and (b) Y-pattern. (c), Energy density as a function of distance from the base in material coordinates. Energy densities are normalized by the energy density $U_{ref}$ of the unsulcified reference state.
}
\label{fig3}
\end{figure}

To understand the patterns qualitatively, we note that the formation of a sulcus relaxes stress primarily in the direction perpendicular to it (similar to a crack), so that it is unfavorable for adjacent sulci to be parallel. The arrangement of the I-shaped sulci in a square lattice with alternating orientations is a natural  solution. Indeed, in Fig.\ \ref{fig1} we see such a pattern although  the system may get trapped in metastable states that break this order at times. For Y-shaped sulci, on the other hand, the simplest plane-filling symmetric pattern is based on a hexagonal lattice, although again we see imperfections at high compression in our unfolding simulations; we note that these symmetries are not those of the underlying lattice \cite{sm}. 

Perfectly ordered sulcus patterns can be constructed numerically by using a ``mold" to imprint the desired pattern onto a precompressed layer. Given the hysteretic nature of the sulficification transition \cite{hohlfeld}, when the mold is removed, the pattern persists. In Fig.\ \ref{fig2}a, we show the square I-pattern in a domain with periodic boundary conditions, simulating a unit cell with two horizontal and two vertical I's on a lattice of dimensions $ \sim 125 \times 125 \times 40$ with a simulation domain of unknown side length $a_s$. For each simulation we fix $\lambda$ and vary $a_s$ to find the energy minimum. In Fig.\ \ref{fig2}b, we show the hexagonal Y-pattern  and again find the optimal hexagonal spacing $a_h$ for a given fixed compression. The surface profile of an I-sulcus shown in Fig.\ \ref{fig2}a reveals that it is symmetric while the surface profile of a Y-sulcus (Fig.\ \ref{fig2}b) reveals that it has a sharp asymmetric center. A comparison of the energies of these patterns with those of freely relaxed layers in Fig.\ \ref{fig2}c shows that the perfectly ordered patterns have lower energy. Furthermore, we find that I-patterns are stable for $\lambda \gtrsim 0.61$ and Y-patterns are stable for $\lambda \lesssim 0.63$, while in the narrow regime $0.61 \le \lambda \le 0.63$  their energies are nearly equal, and explains the coexistence of Y's and I's seen in Fig.\ \ref{fig1}b. The optimal spacing of these structures is given by the relations  $a_s \approx 3h$ and $a_h \approx 2h$ but it increases weakly with compression in both cases.  The square and hexagonal symmetries of the sulcus patterns are similar to those seen in other pattern forming systems such as fluid convection \cite{cross} and elastic fracture \cite{gauthier}; however, there are fundamental differences from a mechanistic perspective, and we will not pursue the mathematical analogies further here. 

Sulcification is an energetic consequence of the exchange of stability between a uniformly deformed state and a set of localized states. To quantify this, in Figs.\ \ref{fig3}a,b we show the depth-averaged energy distributions of the I- and Y-patterns, obtained by integrating energy density over the thickness of the layer in material coordinates. They reveal that Y's span triangular areas on the surface, whereas I's span elliptical areas. Although sulcified states are favored in terms of total energy, between sulci the energy density increases with respect to the reference state. Similarly, energy distribution in the thickness direction (Fig.\ \ref{fig3}c) shows that, while relaxing compression relieves energy near thesurface, the material at some depth gains energy as it conforms to the buckling surface. 

\begin{figure}[t]
\begin{center}
\includegraphics[width = 76mm]{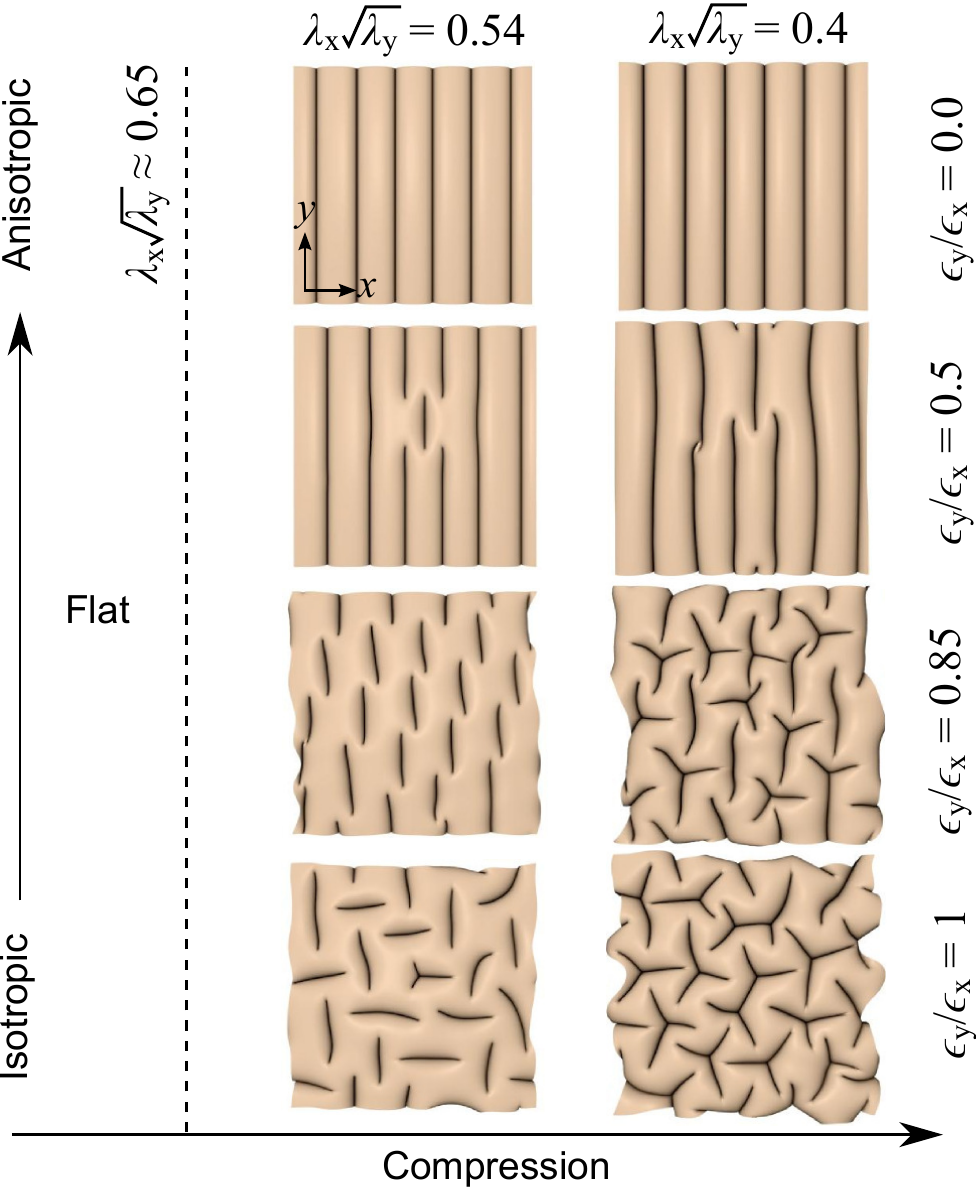}
\end{center}
\caption{Layers with various anisotropic strain ratios $\epsilon_y/\epsilon_x$ (rows) are shown at modest compression ($\lambda_x \sqrt{\lambda_y} = 0.54$, left column) and at high compression ($\lambda_x \sqrt{\lambda_y} = 0.4$, right column).
}
\label{fig4}
\end{figure}

Having considered isotropic compression, we now look at anisotropic compression, characterized by the strain ratio $\epsilon_y/\epsilon_x < 1$ ($\epsilon_x = \lambda_x - 1$, $\epsilon_y = \lambda_y - 1$). We perform these simulations by starting with a compressed reference state and then unloading the layer quasistatically keeping $\epsilon_y / \epsilon_x$ constant. The lattice size and simulation domain are as in the simulation of Fig.\ \ref{fig1}. In the perfectly anisotropic case ($\epsilon_y/\epsilon_x = 0$, see Fig.\ \ref{fig4}) as expected, we see stripes of sulci in the direction perpendicular to the direction of compression. By including compression in the $y$-direction we find that when $\epsilon_y/\epsilon_x \approx 0.5$ stripes begin to break up. When the strains are set almost equal ($\epsilon_y/\epsilon_x = 0.85$ in Fig.\ \ref{fig4}), Y-shaped sulci appear, but anisotropy is still apparent from the pattern. As in the isotropic case ($\epsilon_y/\epsilon_x = 1$), Y's transform to I's with decompression, but now all the I's are oriented perpendicular to the direction of highest compression. We observe an almost identical unfolding threshold $\lambda_x \sqrt{\lambda_y} \approx 0.62$ for all $\epsilon_y / \epsilon_x$; this deviates from previous numerical results for the plane strain case \cite{hohlfeld} where sulci unfold at $\lambda_x \sqrt{\lambda_y} \approx 0.647$ because of the weak dependence of the critical strain on the finite mesh size; here sulci vanish when their size become comparable to the mesh spacing. The near threshold behavior of these sulci calls for a more careful analysis, but the transversely isotropic and plane-strain cases have similar hysteresis effects associated with the presence of two critical points  \cite{hohlfeld}.

\begin{figure}[t]
\begin{center}
\includegraphics[width = 82mm]{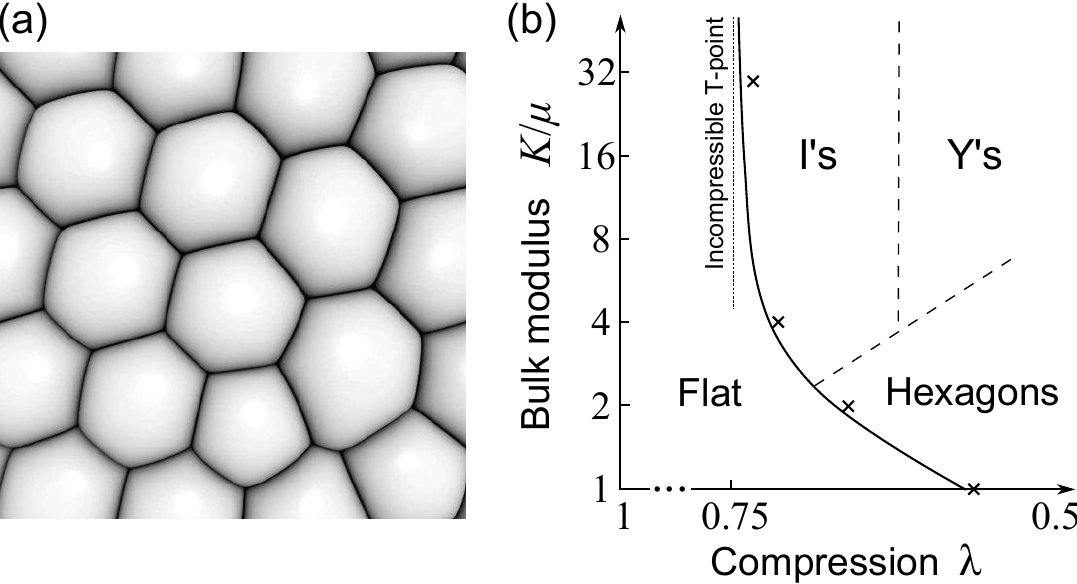}
\end{center}
\caption{(a) A connected sulcus network in a simulated highly compressible solid ($K = 2\mu$, isotropic $\lambda = 0.54$). (b) A schematic diagram summarizing morphologies as a function of bulk modulus and isotropic compression. Sulcification threshold is sketched according to simulated points (crosses). }
\label{fig5}
\end{figure}

Having characterized the patterns on the surface of incompressible hyperelastic solids, we turn briefly to consider the other limit of soft surfaces of highly compressible materials, such as solid foams. Here, the bulk modulus $K \sim \mu$ enters as a relevant parameter, and the Poisson effect  that couples the transverse directions is weaker than in the nearly incompressible materials considered above. When such a solid is isotropically compressed, we find that sulci form a connected hexagonal network, typically with some imperfections, as shown in Fig.\ \ref{fig5}a. For $K \lesssim 2 \mu$ the hexagons persist all the way to the T-point upon unloading, which itself shifts to higher strain with decreasing $K$. These findings can be summarized in a simple phase diagram (Fig.\ \ref{fig5}b) of sulcus morphologies as a function of bulk modulus and compression. We note that the phase boundaries in the diagram are only qualitative guides since the simulations indicate regions of coexistence of the different morphological states.  

Our simulated patterns are able to capture the range of experimental observations of sulcification in swollen gel layers \cite{trujillo} shown in Figs.\ \ref{fig1}e and \ref{fig1}f. Indeed, prior observations show the domains of ordered I patterns and YI-mixtures, reminiscent of our freely relaxed layers, although the transition from I's to Y's has not been previously attributed directly to increasing compression. Furthermore, the spacing between sulci in our simulations agrees well with the experimentally observed spacing \cite{trujillo} with a similar weak strain dependence as observed for uniaxially compressed sulci \cite{cai}. Sulcus patterns in compressible hydrogels (which are poroelastic and thus compressible over long time scales) have been observed to relax with time into honeycomb structures \cite{tanaka, suematsu} that are similar to our compressible hexagon patterns.  In a biological setting, several organs, including the cerebral cortex and cerebellum in the brain \cite{welker}, have sulcified surfaces, with the cerebellum showing striped patterns while the cortex showing triple-junctions that are similar to those seen in our simulations. Recent experiments \cite{xu} show the presence of residual strains in these tissues, consistent with the hypothesis that sulicfication might be a simple consequence of relative growth. From a technological perspective, since sulci form so easily on the surface of soft solids,  they should be easy to manipulate as well. Efforts to design and control smart surfaces using temperature-responsive gels \cite{kim}, voltage-responsive elastomers \cite{wang} and mechanical strain are just beginning, and point the way to functional patterning via sulcification.

We thank E.\ Hohlfeld and R.\ C.\ Hayward for discussions. TT acknowledges the Academy of Finland for funding. The computational resources were provided by the Finnish IT Center for Science (CSC).

\newpage
\onecolumngrid

\begingroup  
  \centering
  \LARGE Supplementary material for `Surface sulci in squeezed soft solids'\\ [1em]
   \endgroup
\begingroup
\centering
T. Tallinen, J. S. Biggins \& L. Mahadevan

\endgroup

\section{Mesh geometry}
\begin{figure}[h]
\centering
\includegraphics[width=120mm]{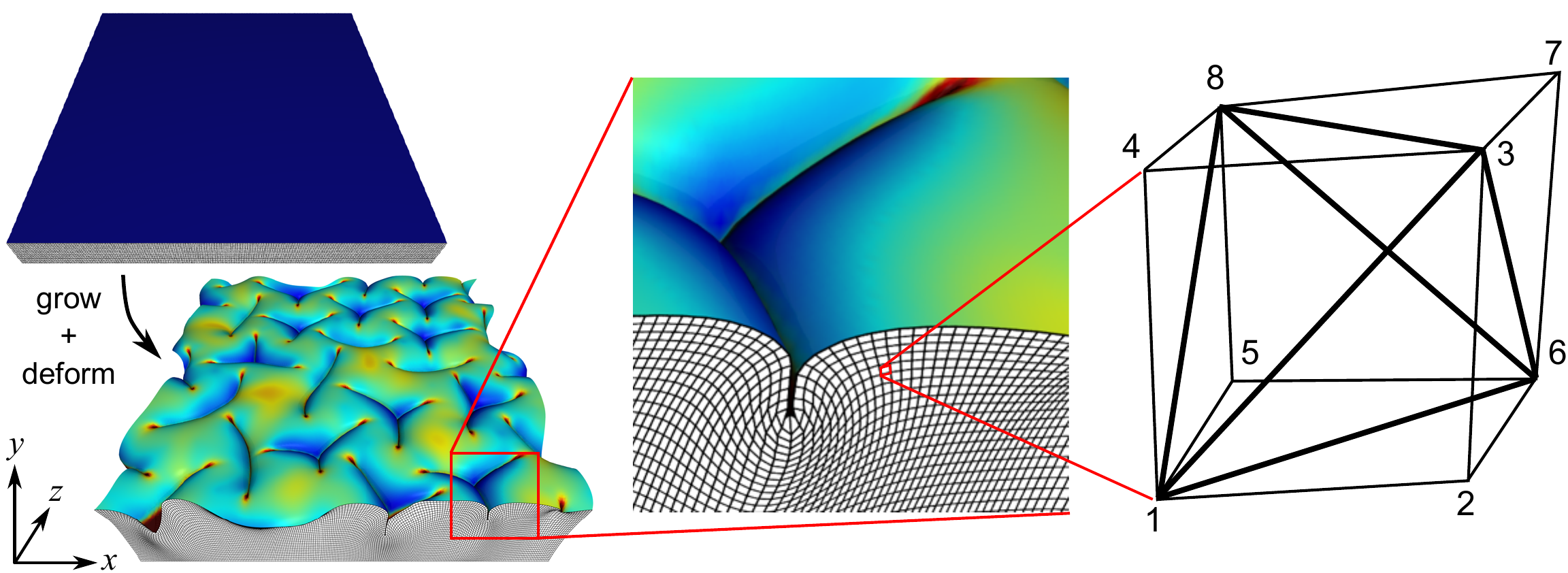}
\caption{An initially flat layer, constructed as a rectangular mesh of tetrahedrons, grows biaxially and relaxes. One rectangle is shown in detail, from which a tetrahedron formed by nodes (1, 3, 6, 8) is indicated by bold lines.}
\label{mesh}
\end{figure}

A layer with thickness $h$ in the stress-free state is confined to a square domain of width $W$. The surface of the layer is assumed to be normal to the $y$-direction and its base is clamped. We assume periodic boundary conditions along the the edges in the $x$ and $z$-directions. The layer is discretized into a rectangular mesh, and each rectangle is divided into five tetrahedrons as indicated in Fig.\ \ref{mesh}. The arrangement of the tetrahedrons in any two neighboring rectangles is reflected with respect to the face they share; this imposes mesh symmetry with respect to reflections in the $x$, $y$ and $z$-directions. The number of nodes in the planar and vertical directions is adjusted so that each edge of the rectangle has  length $a$.

\section{Tetrahedron element}

\begin{figure}[t]
\centering
\includegraphics[width=85mm]{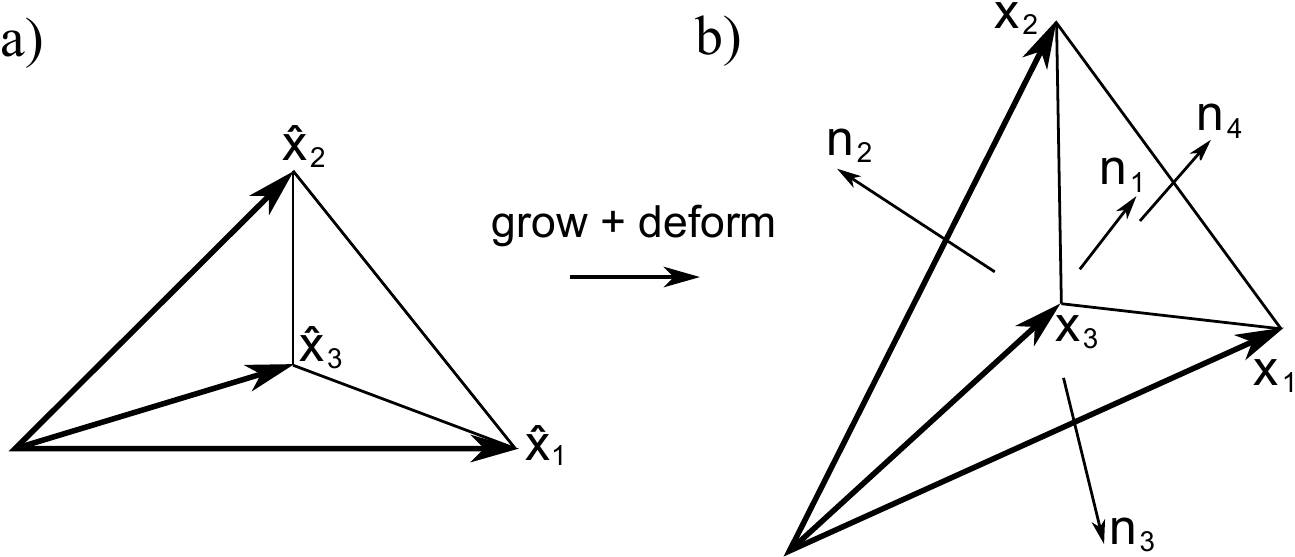}
\caption{A tetrahedron in its (a) stress-free configuration and (b) grown and deformed configuration.}
\label{tetra}
\end{figure}

In its  stress-free configuration, each tetrahedron is defined by its natural state in terms of the matrix
\begin{equation}
\hat{\mathbf{A}} = [\hat{\mathbf{x}}_1 \quad \hat{\mathbf{x}}_2 \quad \hat{\mathbf{x}}_3],
\end{equation}
where $\hat{\mathbf{x}}_1$, $\hat{\mathbf{x}}_2$ and $\hat{\mathbf{x}}_3$ are vectors describing the tetrahedron, see Fig.\ \ref{tetra}a. 
The growth, or swelling of the tetrahedral elements is characterized by the tensor
\begin{equation}
\mathbf{G} = 
\begin{pmatrix}
g_x & 0 & 0 \\
0 & g_y & 0 \\
0 & 0 & g_z 
\end{pmatrix},
\end{equation}
where $g_x$, for example, indicates the stress-free growth ratio in the initial $x$-direction. 
Then the deformed configuration (Fig.\ \ref{tetra}b) of the tetrahedron, including growth, is characterized by
\begin{equation} \label{a}
\mathbf{A} = [\mathbf{x}_1 \quad \mathbf{x}_2 \quad \mathbf{x}_3] = \mathbf{F} \mathbf{G} \hat{\mathbf{A}},
\end{equation}
where $\mathbf{x}_1$, $\mathbf{x}_2$ and $\mathbf{x}_3$ are the deformed basis vectors. Here $\mathbf{F}$ is the elastic deformation gradient, and we have assumed a multiplicative decomposition of the total deformation gradient $\mathbf A$ into a form similar to that used in finite strain plasticity or growth processes. This allows us to obtain $\mathbf{F}$ from eq.\ (\ref{a}) by using the relation
\begin{equation}
\mathbf{F} = \mathbf{A} \left( \mathbf{G} \hat{\mathbf{A}} \right)^{-1}.
\end{equation}

To connect the kinematic relations linking deformation and stress, we assume that our elastomer may be modeled as a compressible neo-Hookean material with a strain energy density (in its undeformed configuration)
\begin{equation} \label{W}
W = \frac{\mu}{2} \left[ \textrm{Tr}(\mathbf{F} \mathbf{F}^{\textrm{T}})J^{-2/3} - 3 \right] + K ( J - \log J-1 ),
\end{equation}
where $\mu$ and $K$ are the shear and bulk modulus, respectively, and $J = \textrm{det}(\mathbf{F})$.
The corresponding Cauchy stress, i.e., the force per unit area in the deformed configuration, is given by
\begin{equation} \label{cauchy}
\boldsymbol{\sigma}_{el} = \frac{1}{J}\frac{\partial{W}}{\partial{\mathbf{F}}} \mathbf{F}^{\textrm{T}} = \mu \left[ \mathbf{F} \mathbf{F}^{\textrm{T}}-\frac{1}{3} \textrm{Tr}(\mathbf{F} \mathbf{F}^{\textrm{T}}) \mathbf{I} \right] J^{-5/3} + K \left( 1-\frac{1}{J} \right) \mathbf{I}.
\end{equation}
We note that this allows us to characterize each element in terms of its local constitutive behavior, in contrast with other structural models that use beams, plates or shells as the elementary units constituting the solid. 

Furthermore, for the purposes of computing the equilibrium configurations, we also assume that each element also has internal damping. Since the relative velocities of the nodes are defined by
\begin{equation}
\mathbf{L} = [\mathbf{v}_1 - \mathbf{v}_0  \quad \mathbf{v}_2 - \mathbf{v}_0 \quad \mathbf{v}_3 - \mathbf{v}_0],
\end{equation}
the rate of deformation is given by
\begin{equation}
\mathbf{D} =\frac{1}{2} (\mathbf{L} + \mathbf{L}^{\textrm{T}}) \mathbf{A}^{-1},
\end{equation}
and the viscous stress by
\begin{equation}
\boldsymbol{\sigma}_{\eta} = \eta \mathbf{D}
\end{equation}
where the elemental viscosity is  $\eta$

Then, the traction on each deformed face ($i$ = 1, 2, 3, 4) of the tetrahedron  is given by
\begin{equation}
\mathbf{s}_i = -\boldsymbol{\sigma} \mathbf{n}_i,
\end{equation}
where
\begin{equation}
\boldsymbol{\sigma} = \boldsymbol{\sigma}_{el} + \boldsymbol{\sigma}_{\eta}
\end{equation}
is the total stress and $\mathbf{n}_i$
are normals with lengths proportional to the deformed areas of the faces, see Fig.\ \ref{tetra}b. 
Nodal forces are obtained by distributing the traction of each face equally on to its three vertices.

\section{Self-avoidance of the surface}

Since our computations involve self-contact at the sulci, we must ensure that this is taken care of correctly. As the element faces at the surface of the layer form a lattice of triangles,  self-avoidance is accommodated by processing 1) vertex-triangle contacts and 2) edge-edge contacts. If a separation $d$ between a vertex and triangle is less than the contact range $h = a/3$ it is considered a contact. Contacts are penalized by an energy $K a^2 \left( \frac{h-d}{h} \right)^2$. Contact forces from this potential is interpolated to the nodes of the triangle and an opposite force is given to the vertex. Edge-edge contacts are processed in a similar way. Further details on algorithms for  proximity detection and contact can be found in C.\ Ericson, {\it Real-time collision detection} (Morgan Kaufmann, San Francisco, 2004) and T.\ Tallinen, {\it Numerical studies on membrane crumpling}, Ph.D.\  Thesis, Univ. Jyv\"{a}skyl\"{a} 2009. 

\section{Damped dynamics for energy minimization}
We use damped second order dynamics for energy minimization. After the nodal forces are determined, Newton's equations of motion are solved for the nodes by an explicit scheme,
\begin{align}
\mathbf{v}(t + \Delta t) =& \mathbf{v}(t) + \frac{\mathbf{f(t)} - \gamma \mathbf{v}(t) }{m} \Delta t, \\
\mathbf{x}(t + \Delta t) =& \mathbf{x}(t) + \mathbf{v}(t + \Delta t) \Delta t.
\end{align}
Here $\Delta t = 0.1a/\sqrt{K}$ is the time step, $m = a^3$ mass of a node, and $\gamma = m$ viscous damping. 
Vectors $\mathbf{f}$, $\mathbf{v}$ and $\mathbf{x}$ are force, velocity and position of a node, respectively. 

\section{Alternative discretization: hexagonal mesh}
\begin{figure}[h]
\centering
\includegraphics[width=110mm]{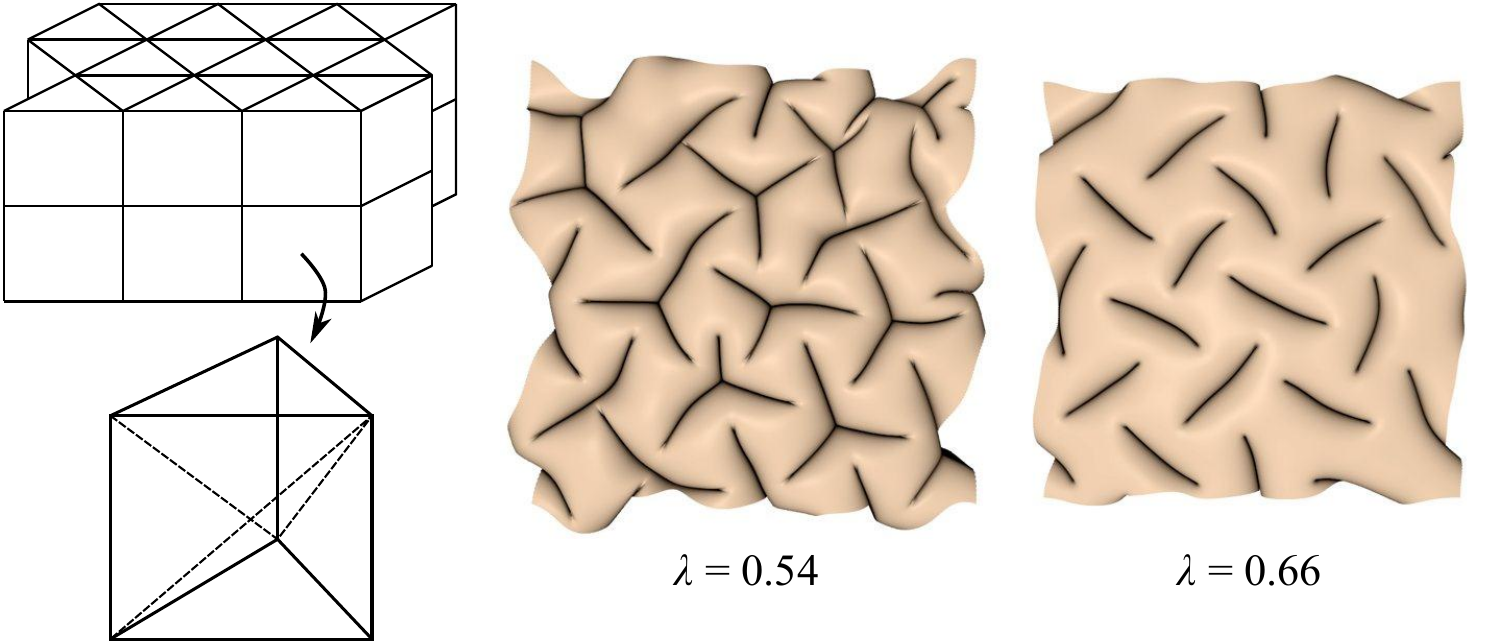}
\caption{A hexagonal prism mesh and division of a prism to three tetrahedrons is illustrated at left. Sulcified layer simulated on the hexagonal mesh (319 $\times$ 371 $\times$ 41 nodes) at two levels of isotropic compression is shown at right. }
\label{hex}
\end{figure}

To explore possible discretization artifacts on the sulcal patterns, we also implemented our simulations  on a hexagonal prism mesh as shown in Fig.\ \ref{hex}. Each prism is divided to three tetrahedrons whose arrengement in any two neighboring prisms is reflected. The reason for this is that division of a prism into tetrahedrons breaks its symmetry, but by the altering arrengement the symmetry is recovered at the level of the whole mesh. We simulate spontaneous sulcification from a compressed reference state followed by quasistatic decompression. We obtain similar patterns , including the transition between Y- and I- phases, as with the rectangular mesh used in the simulations of the main text. The lattice only affects the orientation of the patterns, which simply reflects the fact that in a true continuum system the orientation is arbitrary, while the lattice effects can break this degeneracy to determine the global orientation of the pattern but not its symmetry or morphology.

\begin{figure}[h]
\centering
\includegraphics[width=90mm]{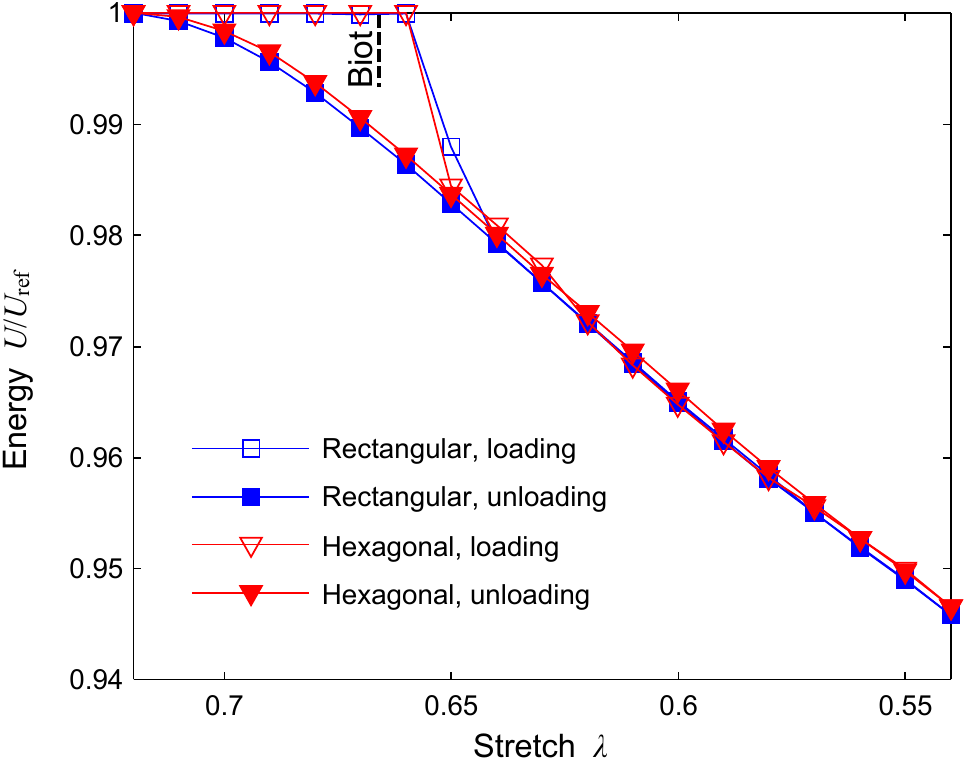}
\caption{A loading/unloading cycle simulated on isotropically compressed rectangular (159 $\times$ 159 $\times$ 41 nodes) and hexagonal (159 $\times$ 185 $\times$ 41 nodes) meshes.}
\label{fullcycle}
\end{figure}

We also simulated full loading/unloading cycles with both meshes to confirm that their energies and thresholds are similar. Figure\ \ref{fullcycle} indicates that under loading the surfaces remain flat, with energy equal to the reference state, until Biot-threshold is reached. Under unloading sulci persisted until $\lambda \approx 0.729$ on a rectangular lattice and $\lambda \approx 0.722$ on a hexagonal lattice. The slightly higher unfolding strain of the hexagonal mesh is most likely due to the mismatch between the mesh geometry and preferred orthogonal arrangement of sulci near the threshold.

\begin{thebibliography}{}

\bibitem{witten}
T.\ A.\ Witten and L.\ M.\ Sander, Phys.\ Rev.\ Lett. {\bf 47}, 1400 (1981).

\bibitem{turing}
A.\ M.\ Turing, Philos.\ Trans.\ R.\ Soc.\ Lond.\ B {\bf 237}, 37 (1952).

\bibitem{euler}
L. Euler, Opera Omnia II, v. 10, 1 (1732).

\bibitem{antman}
S. Antman, {\it Nonlinear problems of elasticity, 2nd edn.}, (Springer, NewYork, 2005) .

\bibitem{griffith}
A.A. Griffith, Philos.\ Trans.\ R.\ Soc.\ Lond.\ A {\bf 221}, 163 (1921).

\bibitem{broberg}
K. Broberg, {\it Cracks and fracture}, (Academic, New York, 1999).

\bibitem{bowden}
N.\ Bowden {\it et al.}, Nature {\bf 393}, 146 (1998).

\bibitem{biot}
M.\ A.\ Biot, {\it Mechanics of incremental deformations} (John Wiley and Sons, New York, 1965).

\bibitem{hohlfeld}
E.\ Hohlfeld and L.\ Mahadevan, Phys.\ Rev.\ Lett. {\bf 106}, 105702 (2011); Phys.\ Rev.\ Lett. {\bf 109}, 025701 (2012). 

\bibitem{suematsu}
N.\ Suematsu, K.\ Sekimoto, and K.\  Kawasaki, Phys.\ Rev.\ A {\bf 41}, 5751 (1990).

\bibitem{tanaka}
T.\ Tanaka {\it et al.}, Nature {\bf 325}, 796 (1987).

\bibitem{dervaux}
J.\ Dervaux, Y.\ Couder, M.\ A.\ Guedeau-Boudeville, and M.\ Ben Amar, Phys.\ Rev.\ Lett. {\bf 107}, 018103 (2011).

\bibitem{jin}
L.\ Jin, S.\ Cai, and Z.\ Suo, Europhys. Lett. {\bf 95}, 64002 (2011).

\bibitem{mora}
S.\ Mora, M.\ Abkarian, H.\ Tabuteau, and Y.\ Pomeau, Soft Matter {\bf 7}, 10612 (2011).

\bibitem{cao}
Y.\ Cao and J.\ W.\ Hutchinson, Proc.\ R.\ Soc.\ A {\bf 468}, 94 (2012).

\bibitem{dervaux_rev}
J.\ Dervaux, M.\ Ben Amar,  Annu.\ Rev.\ Condens.\ Matter Phys. {\bf 3}, 311 (2012).

\bibitem{trujillo}
V.\ Trujillo, J.\ Kim, and R.\ C.\ Hayward, Soft Matter {\bf 4}, 564 (2008).

\bibitem{yoon}
J.\ Yoon, J.\ Kim, and R.\ C.\ Hayward, Soft Matter {\bf 6}, 5807 (2010).

\bibitem{welker}
W.\ Welker, Cerebral Cortex {\bf 8B}, 3 (1989).

\bibitem{xu}
G. Xu {\it et al}., J. Biomech. Eng. {\bf 132}, 071013-1 (2010).

\bibitem{chen}
D.\ Chen, S.\ Cai, Z.\ Suo, and R.\ C.\ Hayward, Phys.\ Rev.\ Lett. {\bf 109}, 038001 (2012).

\bibitem{pkim}
P.\ Kim, M.\ Abkarian, H.\ A.\ Stone, Nat.\ Mater. {\bf 10}, 952 (2011).

\bibitem{sm}
See supplementary material for simulation details and movies.

\bibitem{cross}
M.\ C.\ Cross and P.\ C.\ Hohenberg, Rev.\ Mod.\ Phys. {\bf 65}, 851 (1993).

\bibitem{gauthier}
G.\ Gauthier, V.\ Lazarus, and L.\ Pauchard, Europhys.\ Lett.\ {\bf 89}, 26002 (2010).

\bibitem{cai}
S.\ Cai, D.\ Chen, Z.\ Suo, and R.\ C.\ Hayward, Soft Matter {\bf 8}, 1301 (2012).

\bibitem{kim}
J.\ Kim, J.\ Yoon, and R.\ C.\ Hayward, Nat.\ Mater. {\bf 9}, 159 (2010).

\bibitem{wang}
Q.\ Wang, M.\ Tahir, L.\ Zhang, and X.\ Zhao, Soft Matter {\bf 7}, 6583 (2011).

\end{thebibliography}
\end{document}